\documentclass{PoS}

\title{Lessons from RHIC for the LHC and vice versa}

\ShortTitle{Lessons from RHIC for the LHC and vice versa}

\author{\speaker{Michael J. Tannenbaum}\thanks{Supported by the U.S. Department of Energy, Contract No. DE-AC02-98CH1-886.}\\
Physics Dept., 510c, Brookhaven National Laboratory, Upton, NY 11973-5000,USA\\E-mail: \email{mjt@bnl.gov}}

\abstract{For the past decade, 
measurements of semi-inclusive single
identified particle spectra and two-particle correlations in p-p
and A+A collisions at RHIC have produced a treasure trove of
results which indicate a suppression of hard-scattered partons in
the medium produced in A+A collisions. A suppression $R_{AA}\approx 0.2$ has been measured in the range 
$5\leq p_T\leq 20$ GeV/c 
in central Au+Au collisions at $\sqrt{s_{NN}}=200$ GeV for $\pi^0$~\cite{PXppg080} and surprisingly for single-electrons from the decay of heavy quarks~\cite{PXRAAePRL98}. Both these results have been confirmed at the LHC in central Pb+Pb collisions at $\sqrt{s_{NN}}=2.76$ TeV~\cite{ALICE2010,ALICEDsuppression}. Interestingly, in the $p_T$ range from 5--20 GeV/c the LHC results nearly overlap the RHIC results for $\pi^0$~\cite{PurschkeQM11}. Thus, due to the flatter $p_T$ spectrum, the energy loss in the medium at LHC must be larger than at RHIC in this $p_T$ range. Another issue is whether the (mini) jets from hard-scattering influence the charged particle multiplicity $dN_{\rm ch}/d\eta$ in A+A collisions. This author has long  maintained that this is not possible~\cite{MJT-critical}, which is nicely shown by new data from RHIC at $\sqrt{s_{NN}}=7$ GeV and from LHC~\cite{ALICEmultPRL106} at 2.76 TeV. Also, new at the LHC are the beautiful measurements of the fractional transverse momentum imbalance $A_J$ of di-jets~\cite{ATLASdijetPRL105,CMSPRC} at $\sqrt{s_{NN}}=2.76$ TeV. When corrected~\cite{MJT-Utrecht} for the large fractional imbalance in p-p collisions with the same cuts required to obtain a clean jet sample in Pb+Pb, the relative fractional jet imbalance in Pb+Pb/p-p for $\sim 200$ GeV jets becomes $\approx 15$\%, as confirmed by a new CMS measurement~\cite{CMS-dijet-PLB712}. This imbalance is compared to the same quantity derived at RHIC at $\sqrt{s_{NN}}=200$ GeV from two-particle correlations of fragments from di-jets, using a trigger $\pi^0$ with $p_T\approx 10$ GeV/c. The deduced~\cite{MJT-Utrecht} di-jet fractional imbalance in this lower $p_T$ range and c.m. energy is much larger, $\approx 45$\%. Among other lessons learned from RHIC is the need for p-p and p-A (or d-A) comparison data at the same $\sqrt{s_{NN}}$ in the same detector; and how the heavy-ion results may influence the search for the Higgs particle in p-p collisions at the LHC.    
}

\FullConference{Sixth International Conference on Quarks and Nuclear Physics\\
April 16-20, 2012\\
Ecole Polytechnique,\ Palaiseau,\ Paris}

\def\lsim{\raise0.3ex\hbox{$<$\kern-0.75em\raise-1.1ex\hbox{$\sim$}}}
\def\gsim{\raise0.3ex\hbox{$>$\kern-0.75em\raise-1.1ex\hbox{$\sim$}}}

\def\mean#1{\left<#1\right>}
\def\Journal#1#2#3#4{ {\it{#1}} {\bf #2}, #3 (#4)}

\def\EPJC{{Eur. Phys. J.}\ {\rm C}}

\def\JPG{{J. Phys.}\ {\rm G}}

\def\NPA{{Nucl. Phys.}\ {\rm A}}

\def\PLB{{Phys. Lett.}\ {\rm B}}

\def\PRL{Phys. Rev. Lett.\ }
\def\PRD{{Phys. Rev.}\ {\rm D}}
\def\PRC{{Phys. Rev.}\ {\rm C}}
\def\ARNPS{{Ann. Rev. Nucl. Part. Sci.\ }}

\def\QGP{\Red Q\Blue G\Green P\Black}

\begin{document}

\begin{figure}[h]
\begin{center}
\begin{tabular}{cc}
\includegraphics[width=0.64\linewidth]{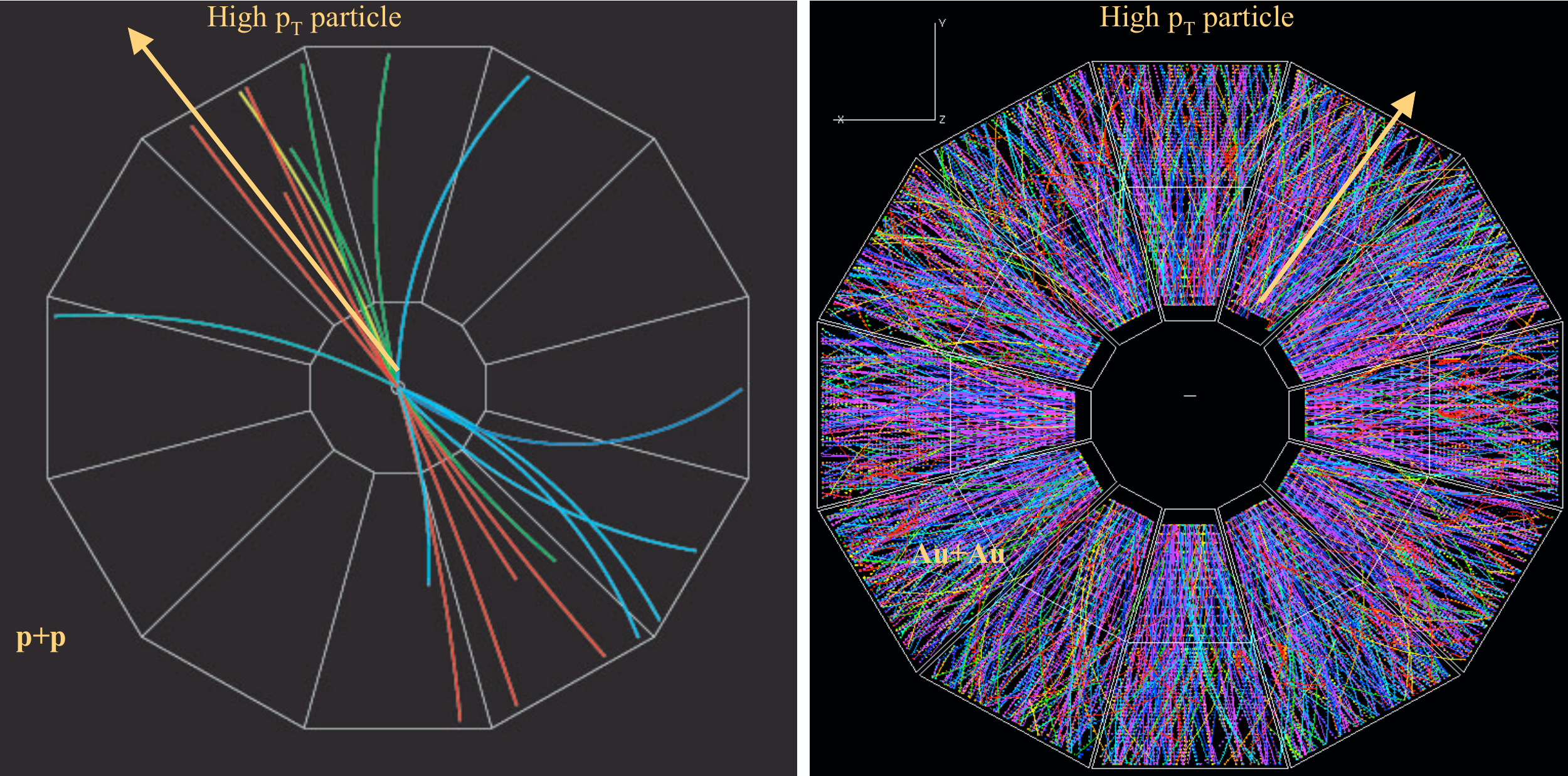}&\hspace*{-0.025\linewidth}
\includegraphics[width=0.315\linewidth,height=0.315\linewidth]{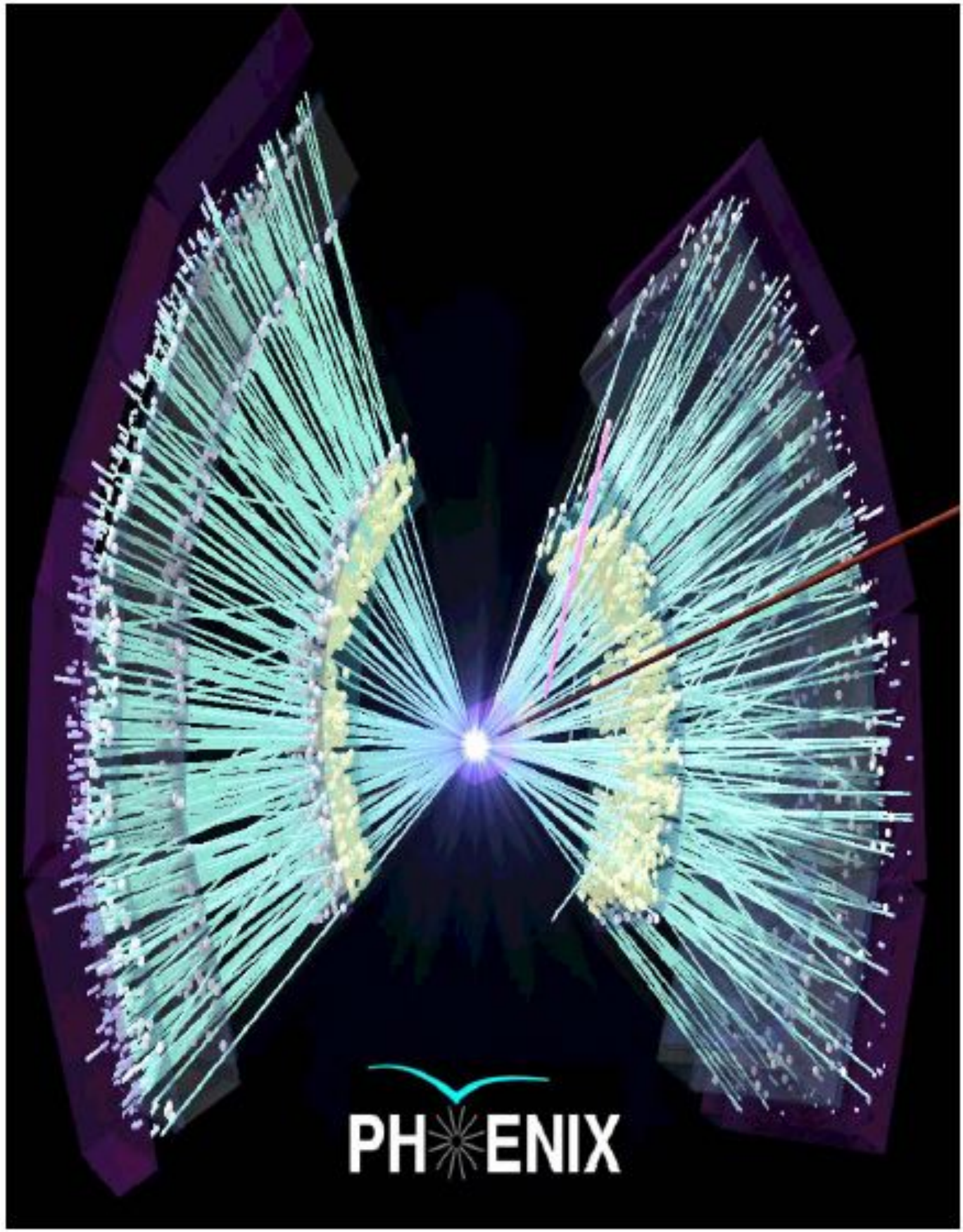}
\end{tabular}
\end{center}\vspace*{-0.15in}
\caption[]{a) (left) A p-p collision in the STAR detector viewed along the collision axis; b) (center) Au+Au central collision at $\sqrt{s_{NN}}=200$ GeV in the STAR detector;  c) (right) Au+Au central collision at $\sqrt{s_{NN}}=200$ GeV in the PHENIX detector.  
\label{fig:collstar}}
\end{figure}

The principal difference in dealing with collisions of relativistic heavy ions, e.g. Au+Au, compared to p-p or e-p (or e-A) collisions at the same nucleon-nucleon c.m. energy, $\sqrt{s_{NN}}$, is that the particle multiplicity is $\sim$ A times larger in A+A central collisions than in p-p collisions as shown in actual events from the STAR and PHENIX detectors at RHIC in Fig.~\ref{fig:collstar}. 

\begin{figure}[!h]
\begin{center}
\begin{tabular}{cc}
\includegraphics[width=0.50\linewidth]{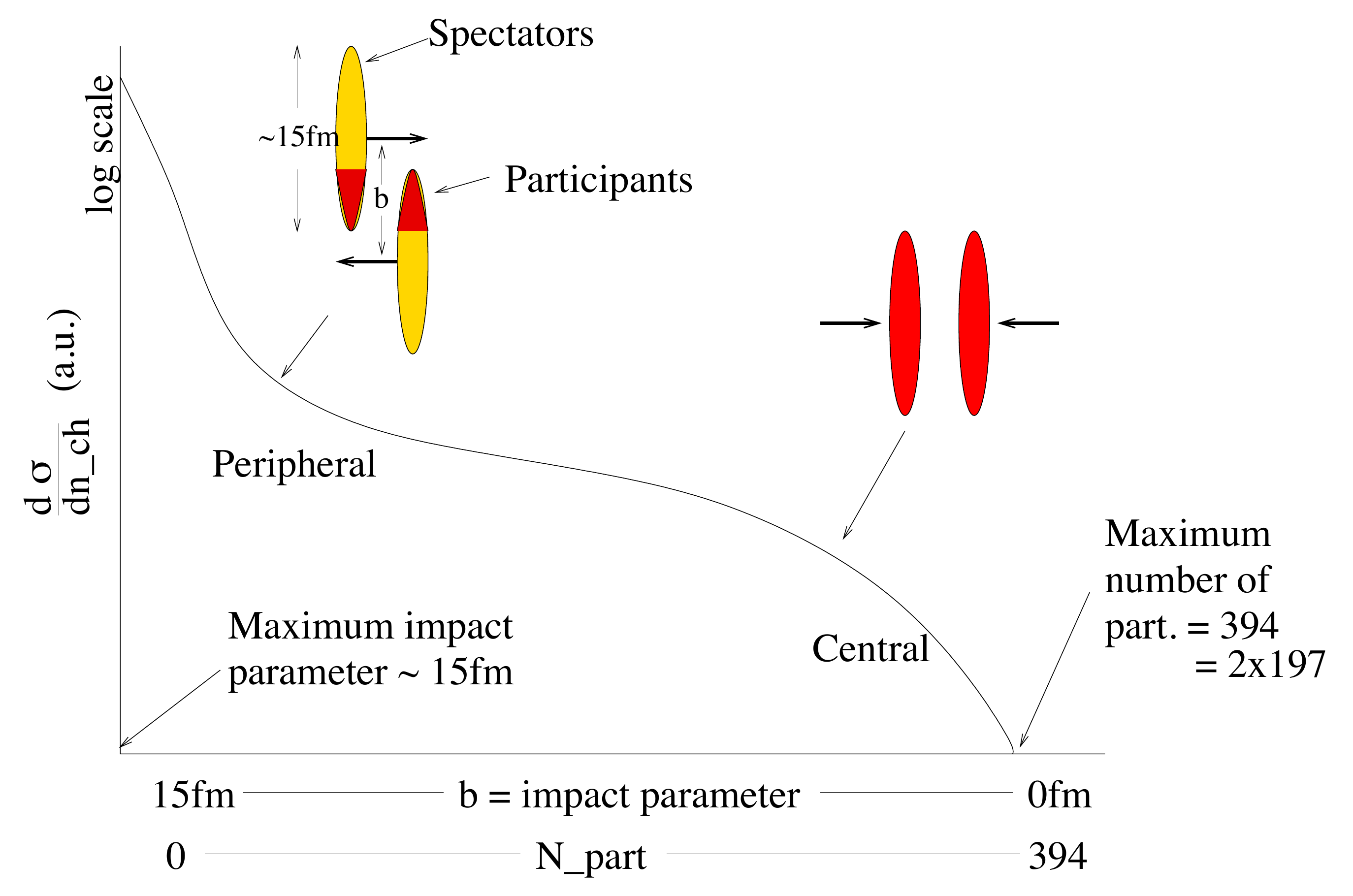}
\includegraphics[width=0.45\linewidth]{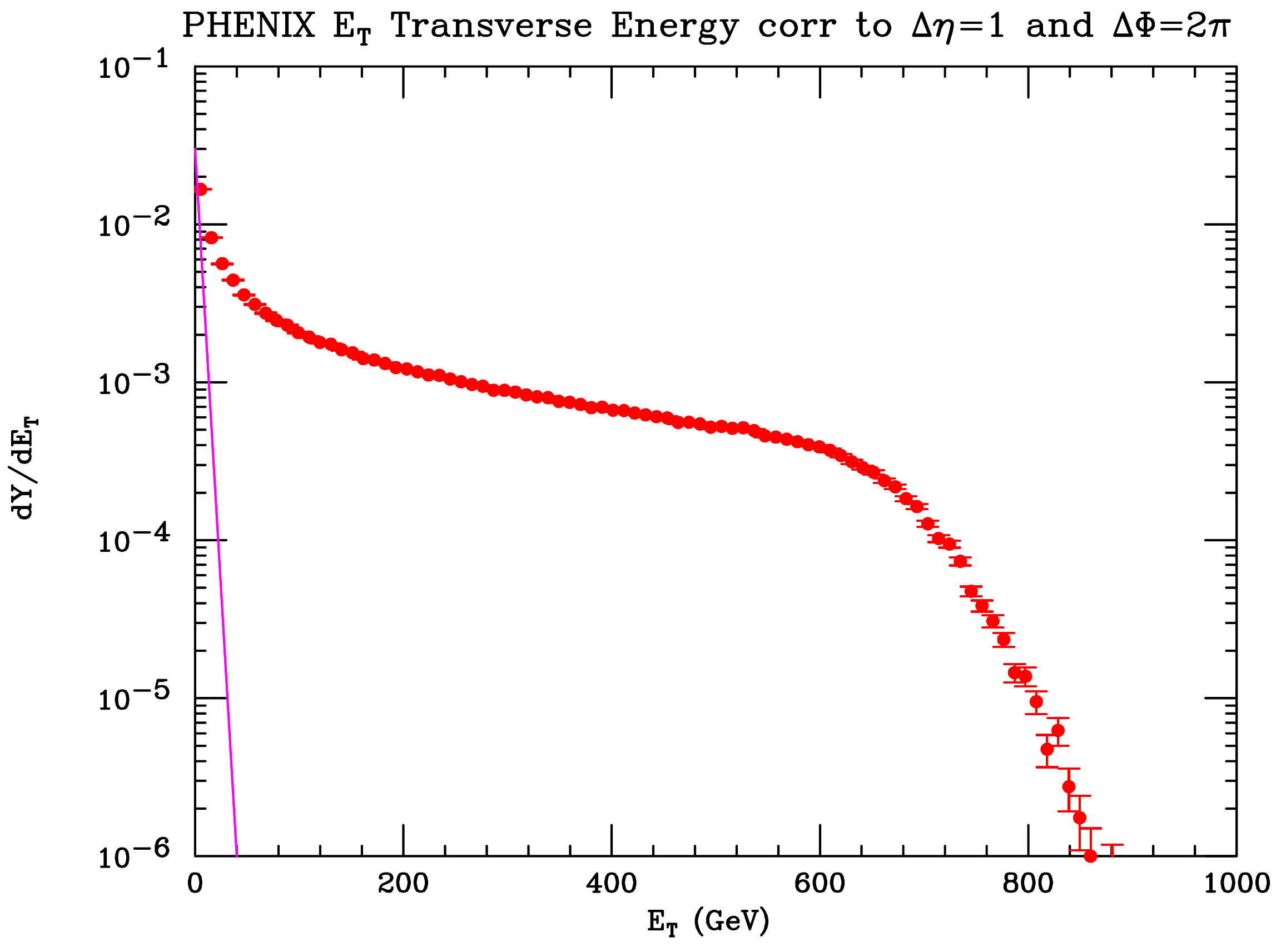}\end{tabular}
\end{center}\vspace*{-0.15in}
\caption[]{a) (left) Schematic of collision of two nuclei with radius $R$ and impact parameter $b$. The curve with the ordinate labeled $d\sigma/d n_{\rm ch}$ represents the relative probability of charged particle  multiplicity $n_{\rm ch}$ which is directly proportional to the number of participating nucleons, $N_{\rm part}$. b)(right) Transverse energy ($E_T=\sum E_i \sin\theta_i$) distribution in Au+Au (data points) and p-p collisions at $\sqrt{s_{NN}}=200$ GeV from PHENIX~\cite{ppg019}.  
\label{fig:nuclcoll}}
\end{figure}

A schematic drawing of a collision of two relativistic Au nuclei is shown in Fig.~\ref{fig:nuclcoll}a. 
In the center of mass system of the nucleus-nucleus collision, the two Lorentz-contracted nuclei of radius $R$ approach each other with impact parameter $b$. In the region of overlap, the ``participating" nucleons interact with each other, while in the non-overlap region, the ``spectator" nucleons simply continue on their original trajectories and can be measured in Zero Degree Calorimeters (ZDC), so that the number of participants can be determined. The degree of overlap is called the centrality of the collision, with $b\sim 0$, being the most central and $b\sim 2R$, the most peripheral. The maximum time of overlap is $\tau_\circ=2R/\gamma\,c$ where $\gamma$ is the Lorentz factor and $c$ is the speed of light in vacuum.  

The energy of the inelastic collision is predominantly dissipated by multiple production of soft particles ($\mean{p_T}\approx 0.36$ GeV/c), where $N_{\rm ch}$, the number of charged particles produced, is directly proportional to the number of participating nucleons ($N_{\rm part}$) as sketched on Fig.~\ref{fig:nuclcoll}a. The impact parameter $b$ can not be measured directly, so the centrality of a collision is defined in terms of the upper percentile of $N_{\rm ch}$ or $E_T$ distributions, e.g. top 10\%-ile, upper $10-20$\%-ile. Unfortunately the ``upper'' and ``-ile'' are usually not mentioned which sometimes confuses the uninitiated.  

\begin{figure}[!h]
\begin{center}
\includegraphics[width=0.55\textwidth]{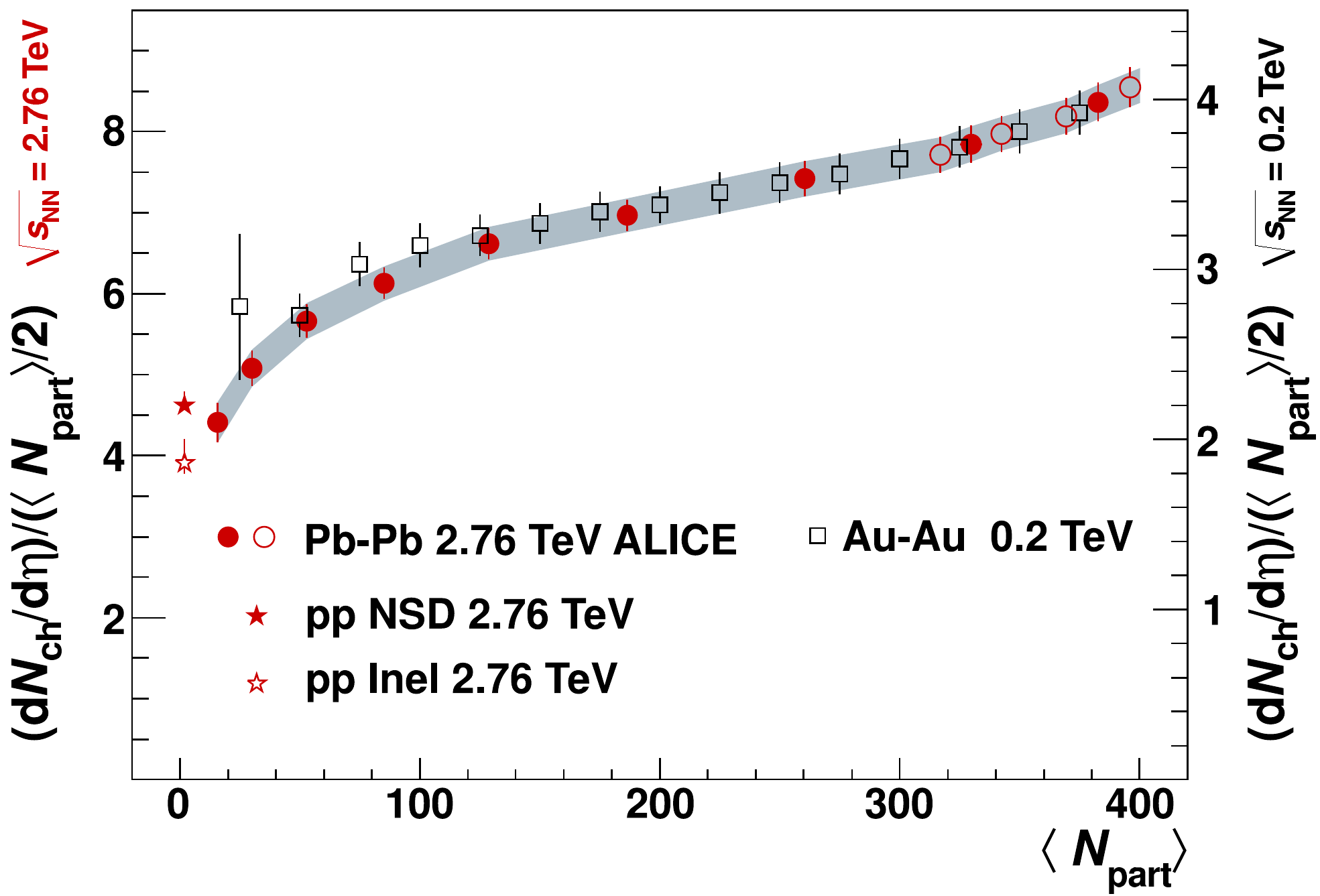}\hspace*{1pc}
\includegraphics[width=0.42\textwidth]{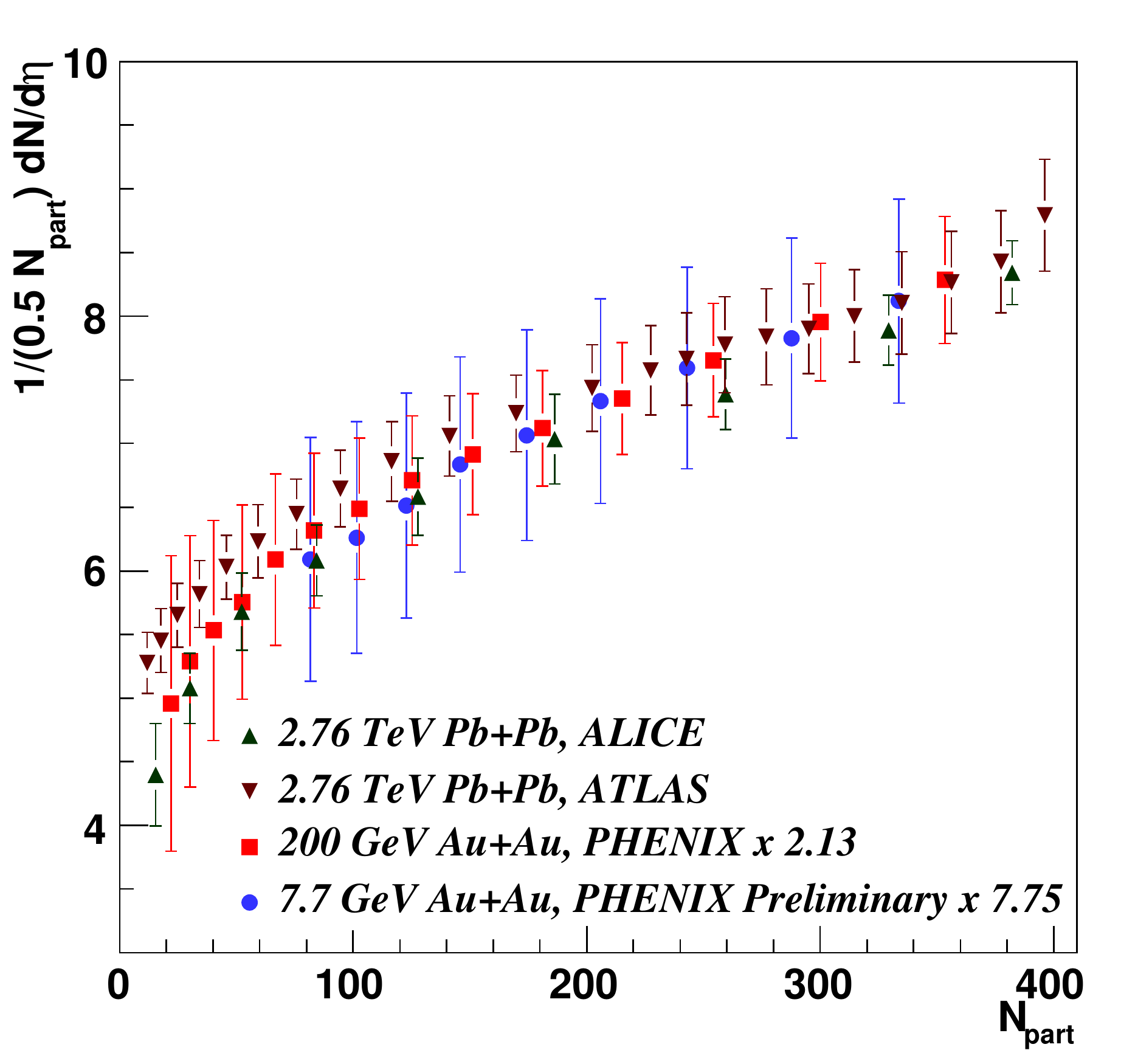}
\end{center}\vspace*{-1pc}
\caption[]{a) (left)Dependence of $(dN_{\rm ch}/d\eta)/(N_{\rm part}/2)$ on the average number of participants $\mean{N_{\rm part}}$   in bins of centrality, for Pb+Pb collisions at $\sqrt{s_{NN}}=2.76$ TeV~\cite{ALICEmultPRL106} and Au+Au collisions at $\sqrt{s_{NN}}=0.200$ TeV~\cite{ppg019}. The scale for the lower-energy data (right side) differs by a factor of 2.1 from the scale for the higher-energy data (left side). b)(right) Data from (a) with new PHENIX measurement in Au+Au at $\sqrt{s_{NN}}=0.0077$ TeV.         \label{fig:dNdeta}}
\end{figure}

In Fig.~\ref{fig:dNdeta}a, measurements of the charged particle multiplicity density $dN_{\rm ch}/d\eta$ at mid-rapidity, $|\eta|<0.5$, relative to the number of participating nucleons, $N_{\rm part}$, are shown as a function of centrality for $\sqrt{s_{NN}}=200$ GeV Au+Au collisions at RHIC~\cite{ppg019} together with new results this year from \mbox{ALICE} in $\sqrt{s_{NN}}=2.76$ TeV Pb+Pb collisions at LHC~\cite{ALICEmultPRL106}. The results are expressed as $(dN_{\rm ch}/d\eta)/(N_{\rm part}/2)$ for easy comparison to p-p collisions (2-participants).  
\vspace*{-0.5pc}

The LHC data show the effect well known from RHIC that $dN_{\rm ch}/d\eta$ does not depend linearly on $N_{\rm part}$, since $(dN_{\rm ch}/d\eta)/(N_{\rm part}/2)$ is not a constant for all $N_{\rm part}$. However the data also show the amazing effect that the ratio of $(dN_{\rm ch}/d\eta)/(N_{\rm part}/2)$ from LHC to RHIC is simply a factor of 2.1 in every  centrality bin.  Thus the LHC and RHIC data lie one on top of each other by simple scaling of the RHIC measurements by a factor of 2.1. In Fig.~\ref{fig:dNdeta}b new results from Au+Au collisions at $\sqrt{s_{NN}}=7.7$ GeV at RHIC scaled by a factor of 7.75 also lie on this curve. The identical shape of the centrality dependence of charged particle production for all $\sqrt{s_{NN}}$ indicates that the dominant effect is the nuclear geometry of the A+A collision. It has been shown that the geometry represents the number of constituent-quark participants/nucleon participant~\cite{AQM}. 

    One of the most important lessons from RHIC is the use of hard-scattering as an in-situ probe of the medium produced in A+A collisions by the effect of the medium on outgoing hard-scattered  
\centerline{\mbox{\includegraphics[width=0.7\textwidth]{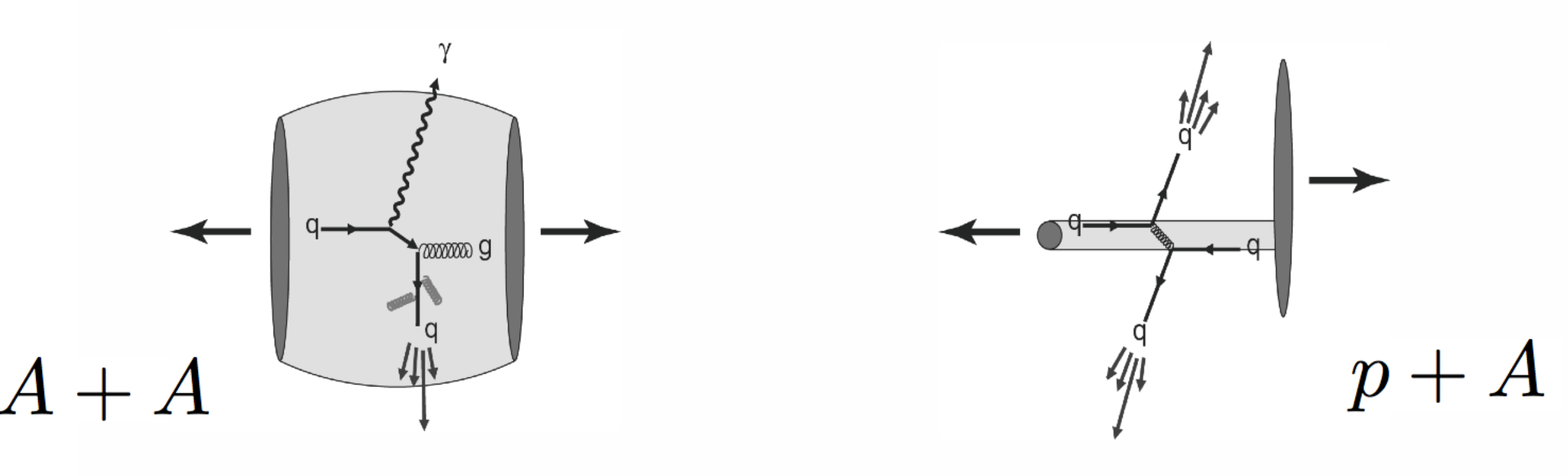} }}  
partons produced by the initial A+A collision. Measurements in p+A (or d+A) collisions, where no (or negligible) medium is produced, allow correction for any modification of the nuclear structure function from an incoherent superposition of proton and neutron structure functions. 

\begin{figure}[!thb]
\begin{center}
\begin{tabular}{cc}
\includegraphics[width=0.46\linewidth]{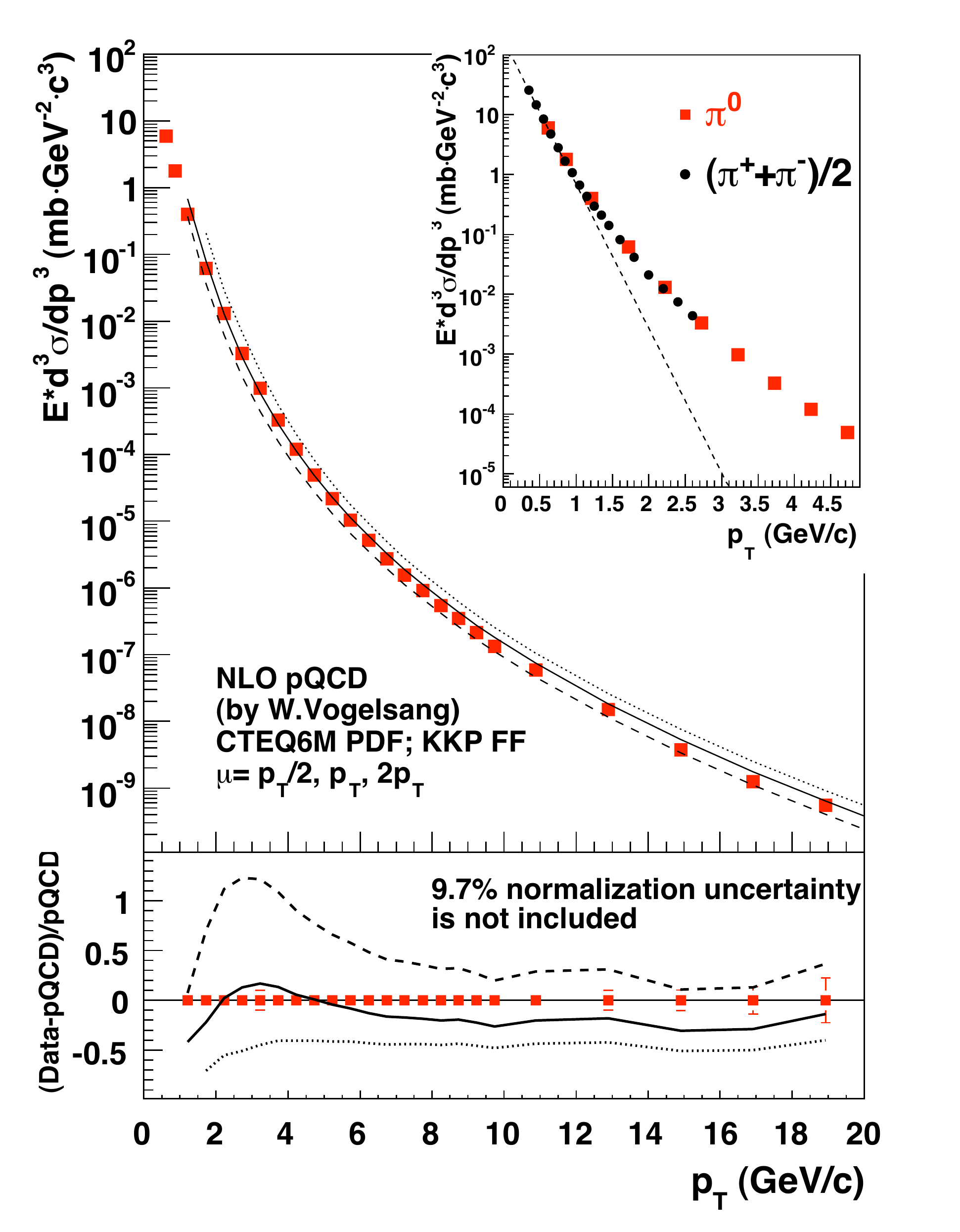}&
\hspace*{-0.04\linewidth}
\includegraphics[height=0.48\linewidth]{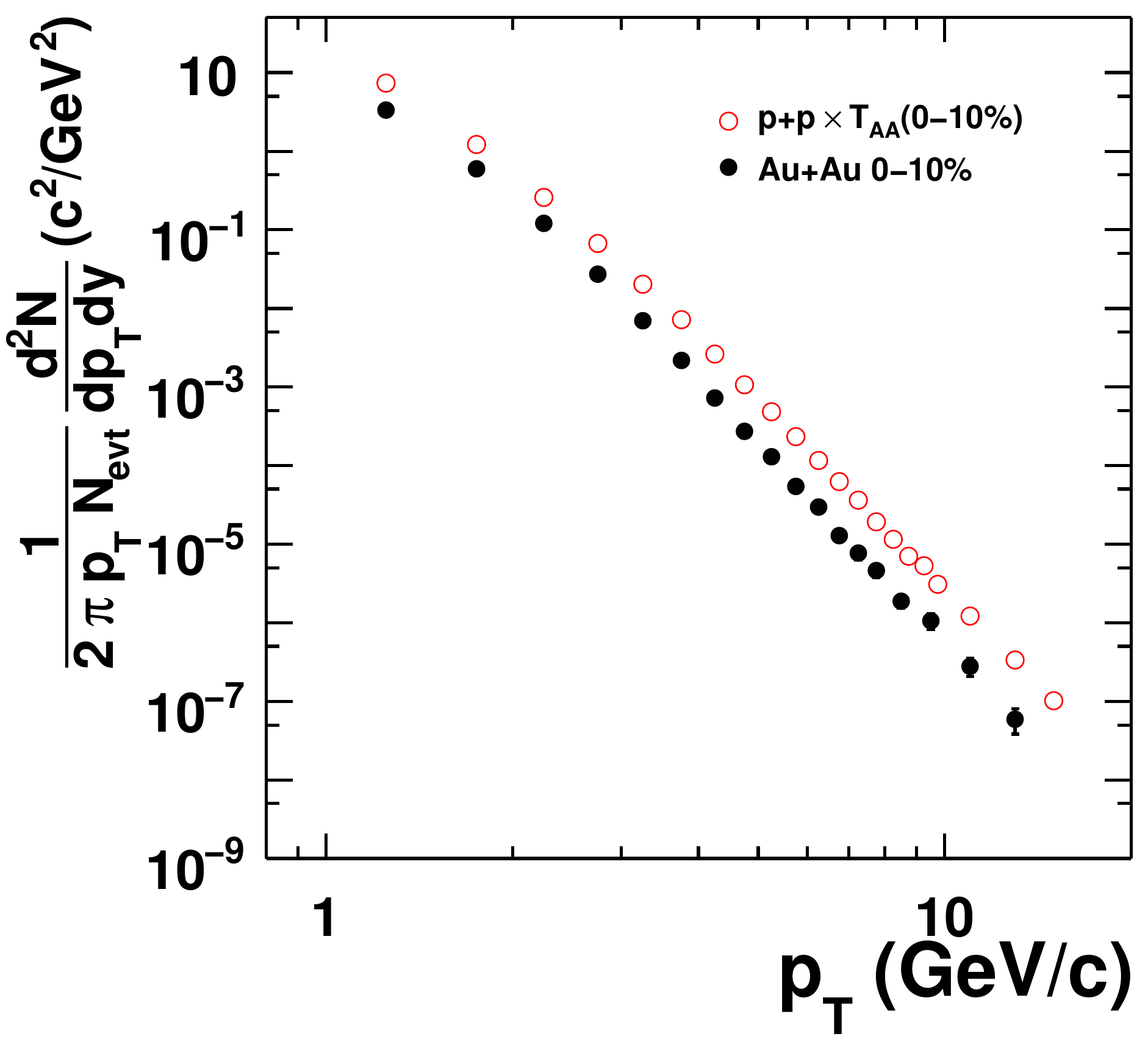}
\end{tabular}
\end{center}\vspace*{-0.25in}
\caption[]{(left) PHENIX measurement of invariant cross section, $E {d^3\sigma}/{d^3p}$, as a function of transverse momentum $p_T$ for $\pi^0$ production at mid-rapidity in p-p collisions at c.m. energy $\sqrt{s}=200$~GeV~\cite{ppg063}. (right) Log-Log plot of the invariant cross section  of $\pi^0$ at $\sqrt{s_{NN}}=200$ GeV as a function of transverse momentum $p_T$ in p-p collisions, multiplied by the $\mean{T_{AA}}$ for Au+Au central (0--10\%) collisions, compared to the measurement~\cite{PXpi0PRC76} of the invariant yield of $\pi^0$ per Au+Au central (0--10\%) collision.  }
\label{fig:Tshirt}
\end{figure}

    The discovery, at RHIC~\cite{PXpi0discovery}, that $\pi^0$ with $p_T\geq 3$ GeV/c are suppressed in central Au+Au collisions is arguably {\em the}  major discovery in Relativistic Heavy Ion Physics. In p-p collisions  at $\sqrt{s_{NN}}=200$ GeV (Fig.~\ref{fig:Tshirt}a)~\cite{ppg063}, the production of $\pi^0$'s via hard-scattering of the quark and gluon constituents of the incident nucleons is indicated by the break, from an exponential dependence of the invariant cross section, $E d^3\sigma/dp^3$, at low $p_T (\leq 1$ GeV/c), to a power-law behavior $p_T^{-n}$ for $p_T\geq 3$ GeV/c, with $n=8.1\pm 0.05$. 
In Au+Au central collisions (Fig~\ref{fig:Tshirt}b)~\cite{PXpi0PRC76}, the suppression is indicated by comparison of the measured yield of $\pi^0$ to the point-like scaled p-p cross section. 

\begin{figure}[!ht]
\begin{center}
\begin{tabular}{cc}
\raisebox{0.01\textheight}{\includegraphics[width=0.51\linewidth,height=0.51\linewidth]{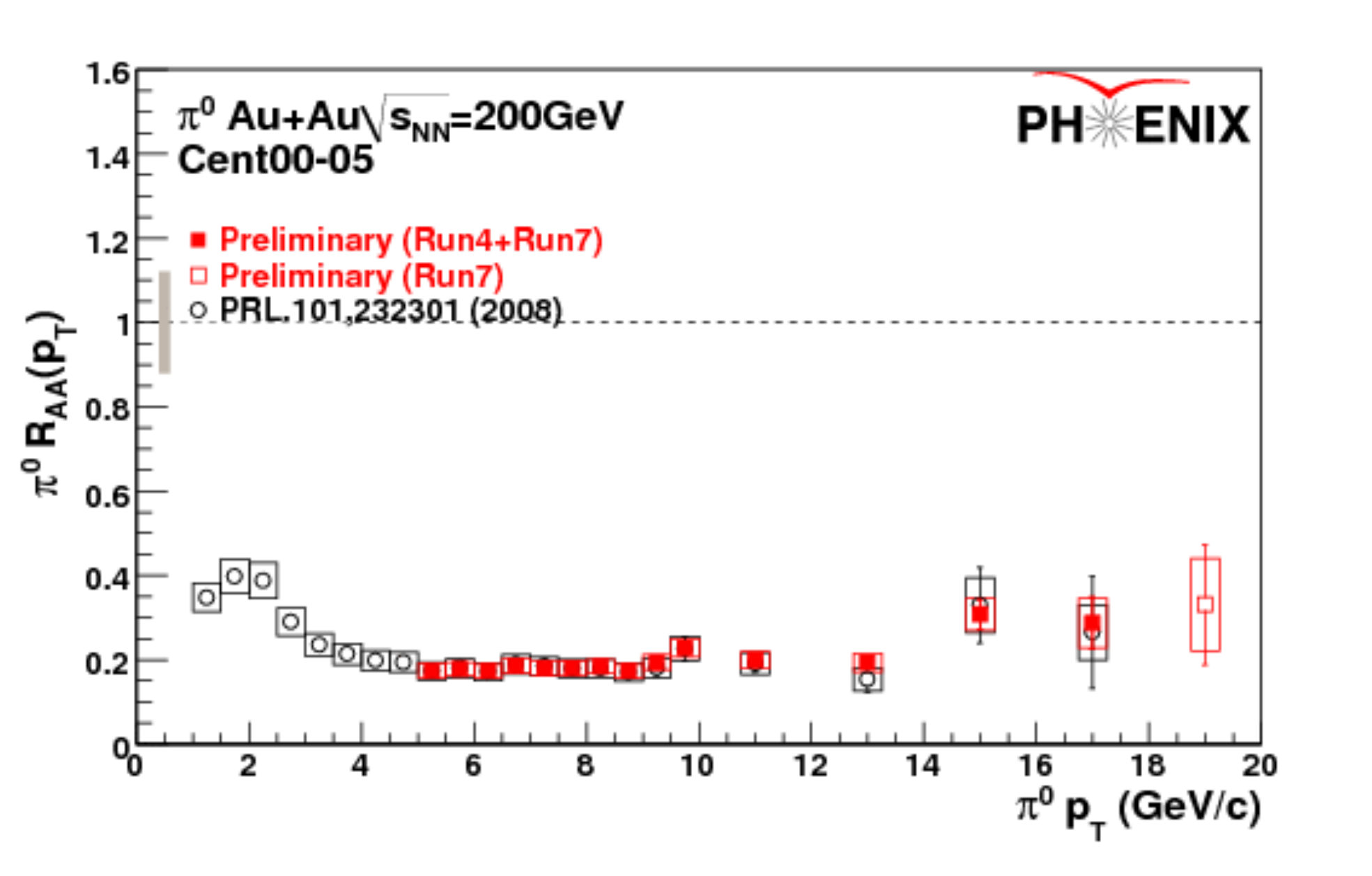}}&
\includegraphics[height=0.50\linewidth]{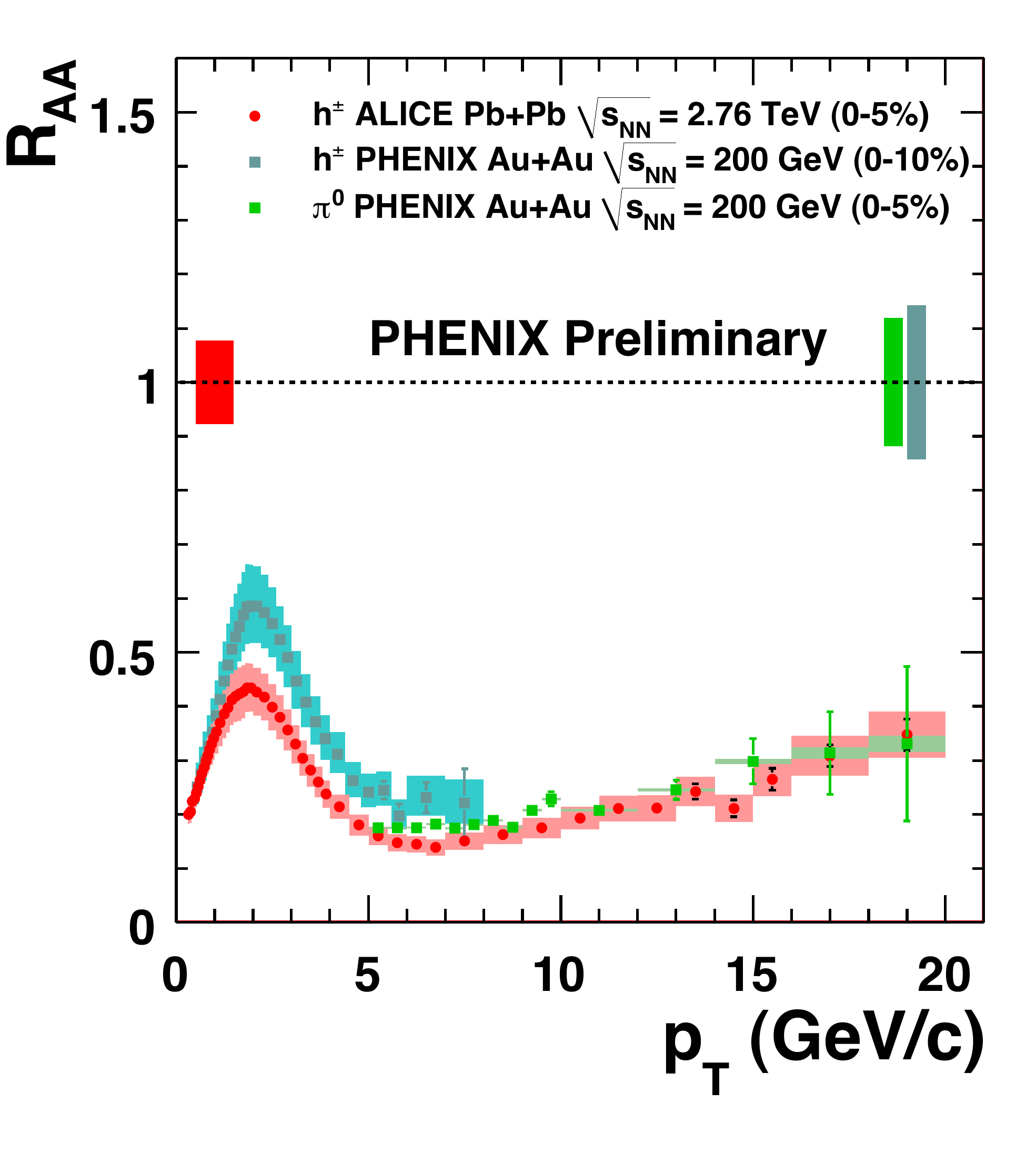}
\end{tabular}
\end{center}\vspace*{-0.25in}
\caption[]{a) (left) PHENIX published~\cite{PXppg080} and preliminary~\cite{PurschkeQM11}  measurements of $R_{AA}$ of $\pi^0$ as a function of $p_T$ in central (0-5\%) Au+Au collisions at $\sqrt{s_{NN}}=200$ GeV. b) ALICE measurements~\cite{ALICE2010} of $R_{AA}(p_T)$ for non-identified charged particles ($h^{\pm}$) in central (0-5\%) Pb+Pb collisions at $\sqrt{s_{NN}}=2.76$ TeV compared to PHENIX measurements of $\pi^0$ from (a) and $h^{\pm}$~\cite{PurschkeQM11}. }
\label{fig:RAA-PXvsALICE}
\end{figure}

Formally, the suppression in central Au+Au collisions at $\sqrt{s_{NN}}=200$ GeV (Fig.~\ref{fig:RAA-PXvsALICE}a)~\cite{PurschkeQM11} 
is presented as the Nuclear Modification Factor, $R_{AA}(p_T)$, the ratio of the yield of $\pi^0$'s  per central Au+Au collision (upper 5\%-ile of observed multiplicity)  to the point-like-scaled p-p cross section:
   \begin{equation}
  R_{AA}(p_T)=[{{d^2N^{\pi}_{AA}/dp_T dy N_{AA}}]/ [{\langle T_{AA}\rangle d^2\sigma^{\pi}_{pp}/dp_T dy}}] \quad , 
  \label{eq:RAA}
  \end{equation}
where $\mean{T_{AA}}$ is the overlap integral of the nuclear thickness functions. 
The $R_{AA}(p_T)$ appears to be constant at a value 0.2, for the range $5\leq p_T\leq 14$ GeV/c, with a hint of a reduction of the suppression (increase of $R_{AA}$) for the range $15\leq p_T\leq 20$ GeV/c. Due to the power-law dependence, a constant value of $R_{AA}$ is indicative of a constant fractional energy shift in the $p_T$ spectrum (Fig.~\ref{fig:Tshirt}b) as suggested by the pQCD-based theory~\cite{BSZ} of radiative energy loss of color-charged partons passing through a medium with a ``large density of similarly exposed color charges'' (i.e. a Quark-Gluon Plasma, \QGP) . 

The first measurement from the LHC~\cite{ALICE2010} of suppression of particles from hard-scattering in central Pb+Pb collisions at $\sqrt{s_{NN}}=2.76$ GeV is shown in Fig.~\ref{fig:RAA-PXvsALICE}b. Despite more than a factor of 20 higher $\sqrt{s_{NN}}$, the $R_{AA}$ measurements by ALICE at LHC appear to be nearly identical to those from PHENIX at RHIC for $5\leq p_T\leq 20$ GeV/c, except that for the LHC data, with better statistics, the upward trend of $R_{AA}(p_T)$ over the whole interval is significant. It is interesting to note that due to the flatter $p_T$ spectrum at the LHC ($p_T^n$ with $n\sim 6$ compared to $n=8.1$ at RHIC), the shift in the spectrum in this $p_T$ range (as in Fig.~\ref{fig:Tshirt}b) must be 50\% larger at LHC than at RHIC. 

\begin{figure}[!h]
\begin{center}
\includegraphics[width=0.54\textwidth]{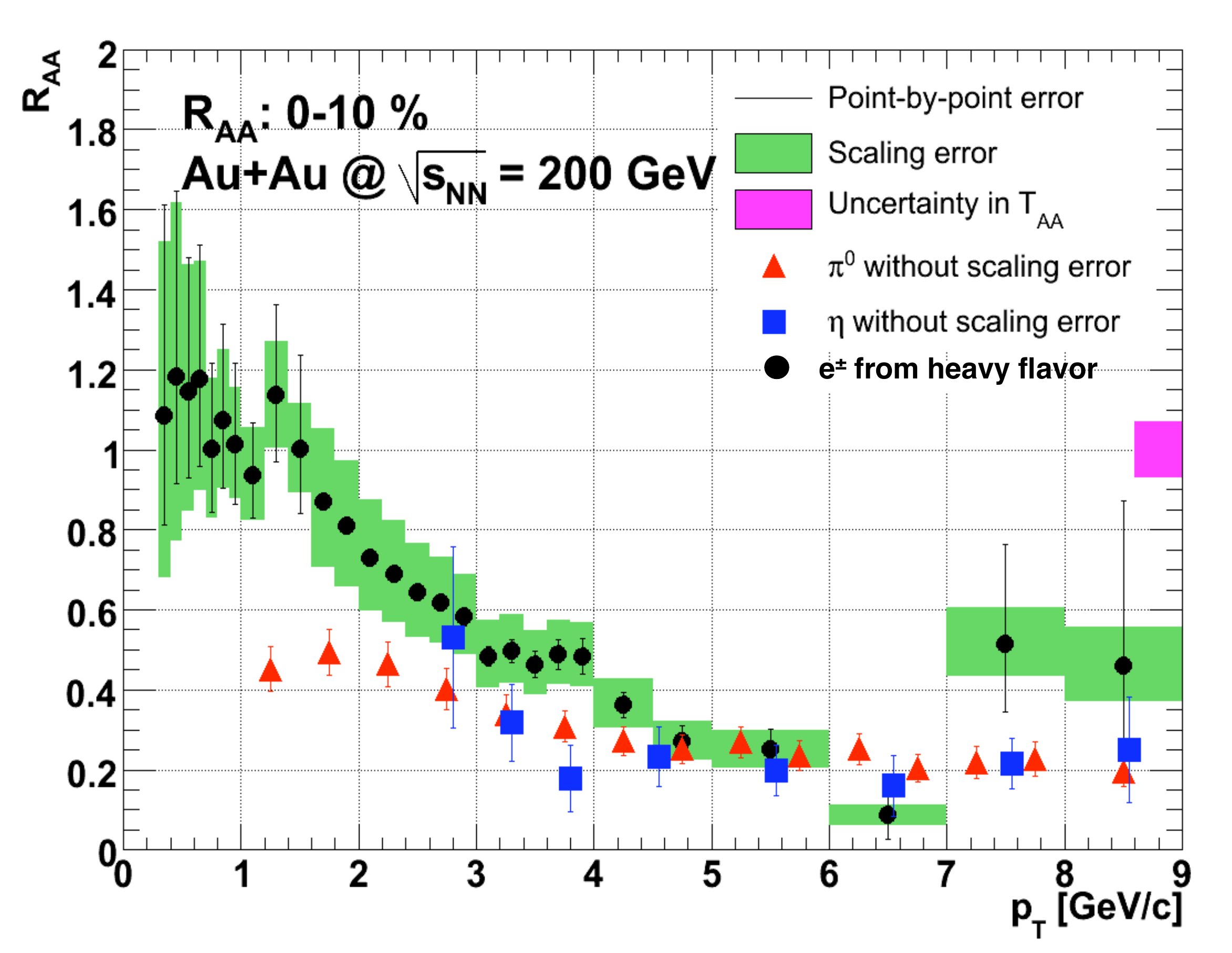}
\includegraphics[width=0.45\textwidth]{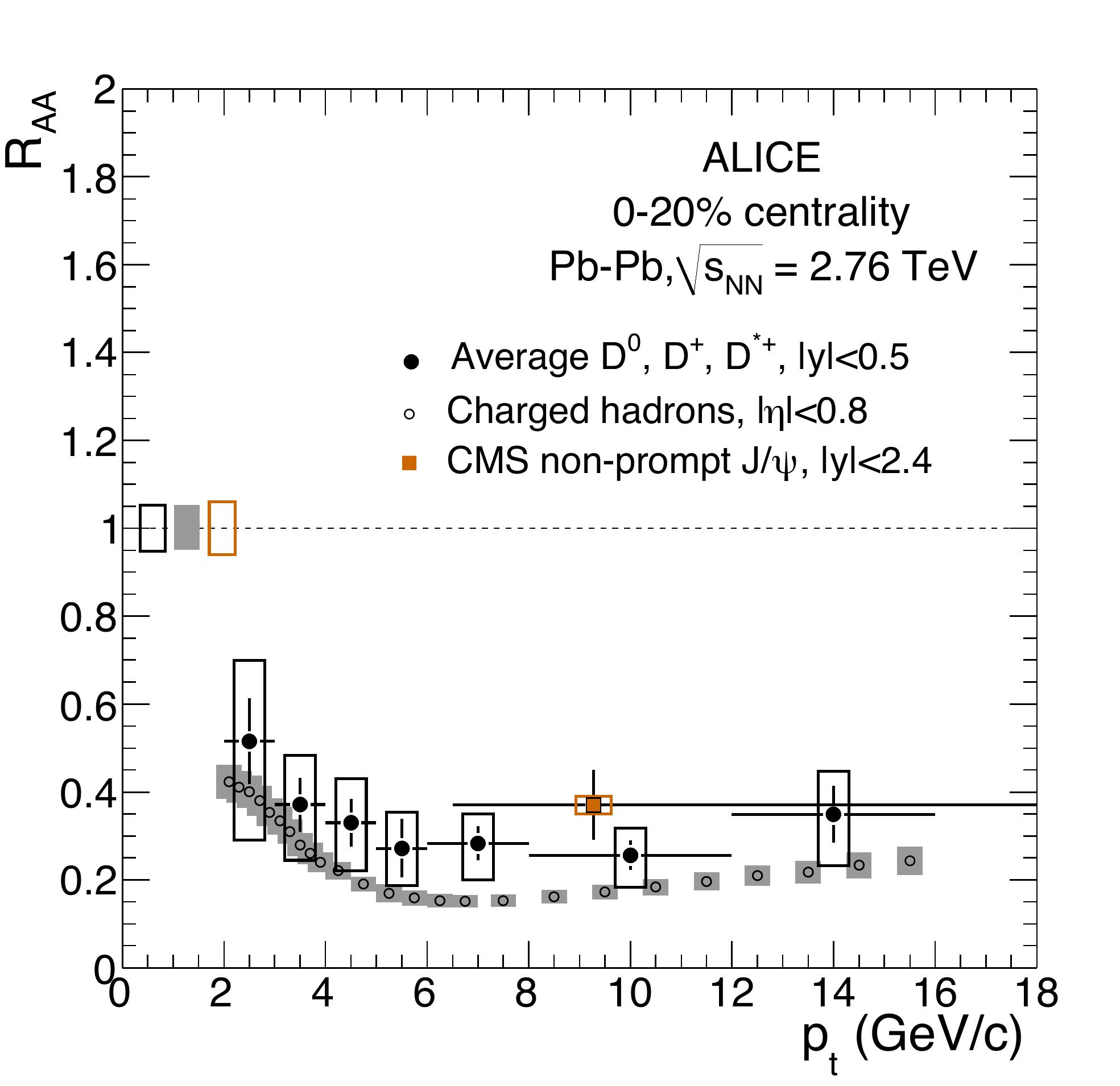}
\end{center}
\caption[]{a) (left) $R_{AA}(p_T)$ measured by PHENIX~\cite{PXRAAePRL98} for direct single-$e^{\pm}$, and $\pi^0$ and $\eta$-mesons in central (0-10\%) Au+Au collisions at $\sqrt{s_{NN}}=200$ GeV. b) (right) $R_{AA}$ of ALICE~\cite{ALICEDsuppression} $D$-mesons, charged hadrons, and CMS~\cite{CMSUps1Ssupp} non-prompt $J/\Psi$, in central (0-20\%) Pb+Pb collisions at $\sqrt{s_{NN}}=2.76$ TeV }
\label{fig:heavy}
\end{figure}

Another discovery at RHIC, now confirmed at the LHC, is the suppression of heavy quarks by the same amount as light quarks for $p_T\gsim 5$ GeV/c. This is indicated at RHIC (Fig.~\ref{fig:heavy}a)~\cite{PXRAAePRL98} by direct single-$e^{\pm}$ from heavy quark ($c$, $b$) decay;   and at the LHC by $D$ mesons from $c$-quarks~\cite{ALICEDsuppression}, and non-prompt $J/\Psi$ from $b$-quarks~\cite{CMSUps1Ssupp} (Fig.~\ref{fig:heavy}b). 
The discovery at RHIC was a total surprise and a problem since it appears to disfavor the radiative energy loss explanation~\cite{BSZ} of suppression (also called jet-quenching) because heavy quarks should radiate much less than light quarks or gluons. 

Many explanations have been offered including some from string theory (see citations in Ref.~\cite{PXRAAePRL98}); but the explanation I prefer was by Nino Zichichi~\cite{AZYukawa} who proposed that since the standard model Higgs Boson,  which gives mass to the Electro-Weak vector Bosons, does not necessarily give mass to Fermions~\cite{Technicolor}, ``it cannot be excluded that in a QCD coloured world (a \QGP), the six quarks are all nearly massless''. If this were true it would certainly explain why light and heavy quarks appear to exhibit the same radiatiative energy loss in the medium. This idea can, in fact, be tested because the energy loss of one hard-scattered parton relative to its partner, e.g. $g+g\rightarrow b+\bar{b}$ , can be measured by experiments at RHIC and LHC using two particle correlations in which both the outgoing $b$ and $\bar{b}$ are identified by measurement of their displaced decay vertices in silicon vertex detectors. When such results are available, they can be compared to $\pi^0$-charged hadron correlations from light quark and gluon jets, for which measurement of the relative energy loss has been demonstrated at RHIC. 

\begin{figure}[!h]
\includegraphics[height=0.22\textheight]{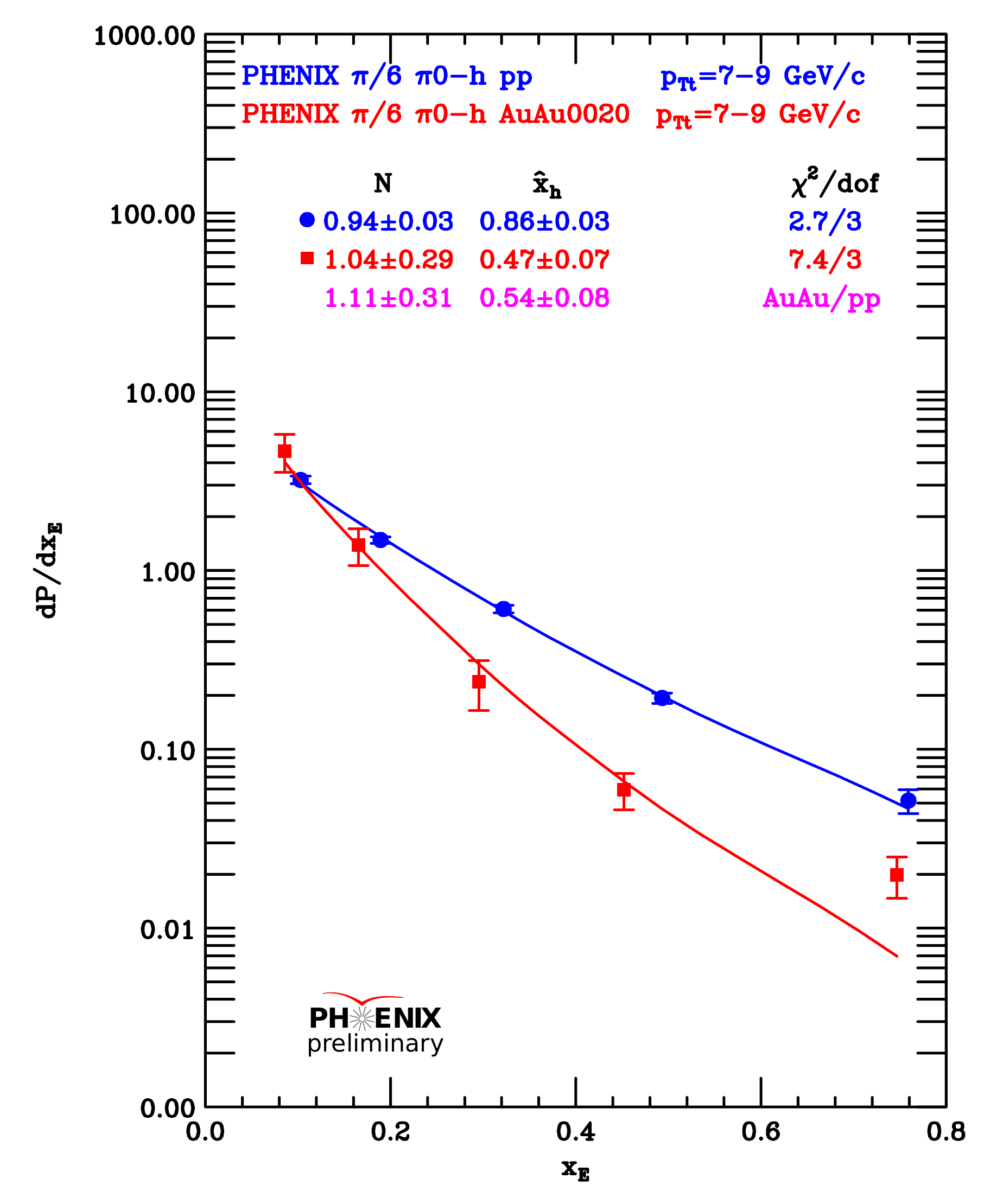}
\includegraphics[height=0.22\textheight]{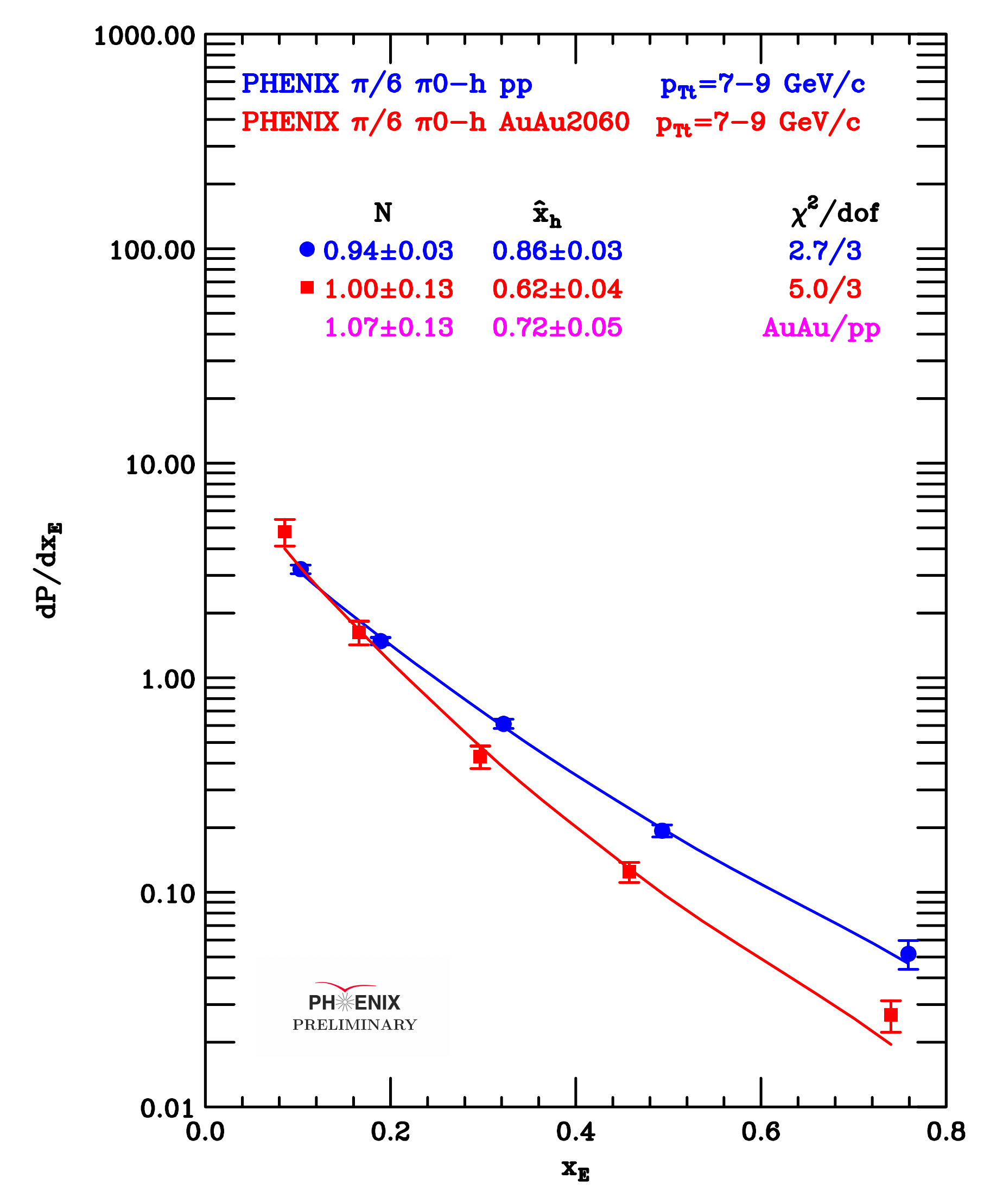}
\includegraphics[height=0.22\textheight]{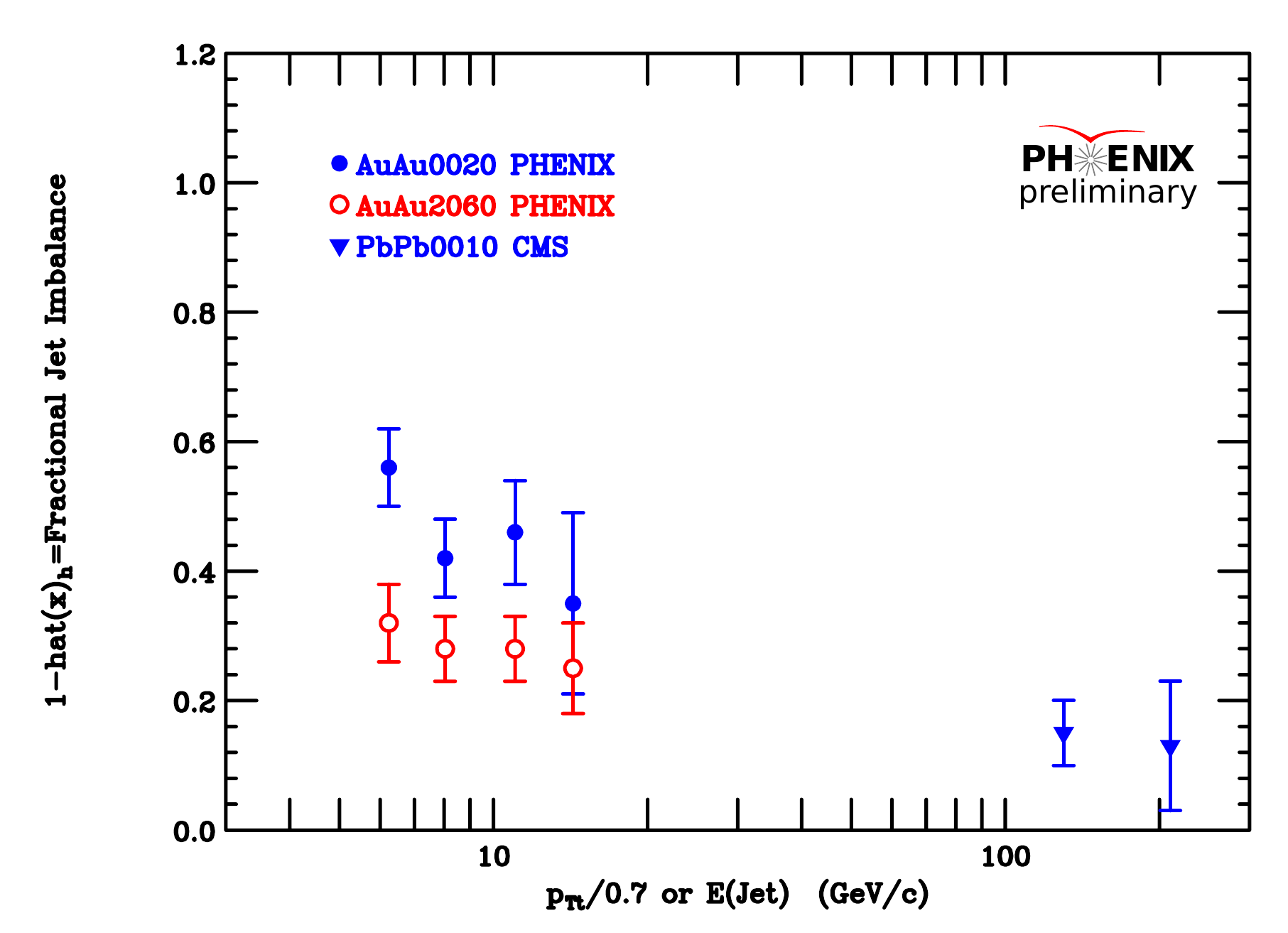}
\caption[]{(left) $x_E$ distributions at RHIC~\cite{MJT-Utrecht} from p-p (blue circles) and AuAu (red squares) collisions for $p_{T_t}=7-9$ GeV/c, together with fits to Eq.~\ref{eq:condxeN2} with parameters indicated: p-p (solid blue line) $N^{pp}=0.94\pm0.03$, $\hat{x}_h^{pp}=0.86\pm0.03$. For AuAu (solid red line), the ratios of the fitted parameters for AuAu/pp are also given:   a) 00-20\% centrality, $N^{AA}=1.04\pm0.29$, $\hat{x}_h^{AA}=0.47\pm0.07$, $\hat{x}_h^{AA}/\hat{x}_h^{pp}=0.54\pm0.08$;  b) 20--60\% centrality $N^{AA}=1.00\pm0.13$, $\hat{x}_h^{AA}=0.62\pm0.04$, $\hat{x}_h^{AA}/\hat{x}_h^{pp}=0.72\pm0.05$. c) (right) Fractional jet imbalance~\cite{MJT-Utrecht}, $1-\hat{x}_h^{AA}/\hat{x}_h^{pp}$, for the RHIC data from (a) and (b), and CMS data	~\cite{CMSPRC,MJT-Utrecht}. }
\label{fig:AuAupp79}
\end{figure}

Another important lesson learned at RHIC~\cite{PXppg029PRD74} is that the distribution of particles, with $p_{T_a}$, opposite in azimuth to a trigger particle, e.g. a $\pi^0$ with large $p_{T_t}$, which is itself the fragment of a jet, does not measure the fragmentation function of the jet opposite in azimuth to the trigger, but, instead, measures the ratio of $\hat{p}_{T_a}$, of the away-parton, to $\hat{p}_{T_t}$, of the trigger-parton, and depends only on the same power $n$ as the invariant single particle spectrum:  \vspace*{-0.5pc}
\begin{equation}
{dP / dx_E}|_{p_{T_t}} \approx {N\,(n-1)}/[\,\hat{x}_h 
{(1+ x_E/\hat{x}_h})^{n}] \qquad .\label{eq:condxeN2}\vspace*{-0.5pc}\end{equation}  
This equation gives a simple relationship between the ratio, $x_E\approx p_{T_a}/p_{T_t}$, of the transverse momenta of the away-side particle to the trigger particle, and the ratio of the transverse momenta of the away-jet to the trigger-jet, $\hat{x}_{h}=\hat{p}_{T_a}/\hat{p}_{T_t}$. PHENIX measurements~\cite{PXpi0hPRL104} of the $x_E$ distributions of $\pi^0$-h correlations in p-p and Au+Au collisions at $\sqrt{s_{NN}}=200$ GeV were fit to Eq.~\ref{eq:condxeN2} (Fig.~\ref{fig:AuAupp79}a,b)~\cite{MJT-Utrecht}.
The results for the fitted parameters are shown on the figures. In general the values of $\hat{x}^{pp}_h$ do not equal 1 but vary between $0.8<\hat{x}^{pp}_h<1.0$ due to $k_T$ smearing and the range of $x_E$ covered. In order to take account of the imbalance ($\hat{x}^{pp}_h <1$) observed in the p-p data, the ratio $\hat{x}_h^{AA}/\hat{x}_h^{pp}$ is taken as the measure of the energy of the away jet relative to the trigger jet in A+A compared to p-p collisions. 

\begin{figure}[!h]
\begin{center}
\includegraphics[width=0.70\textwidth]{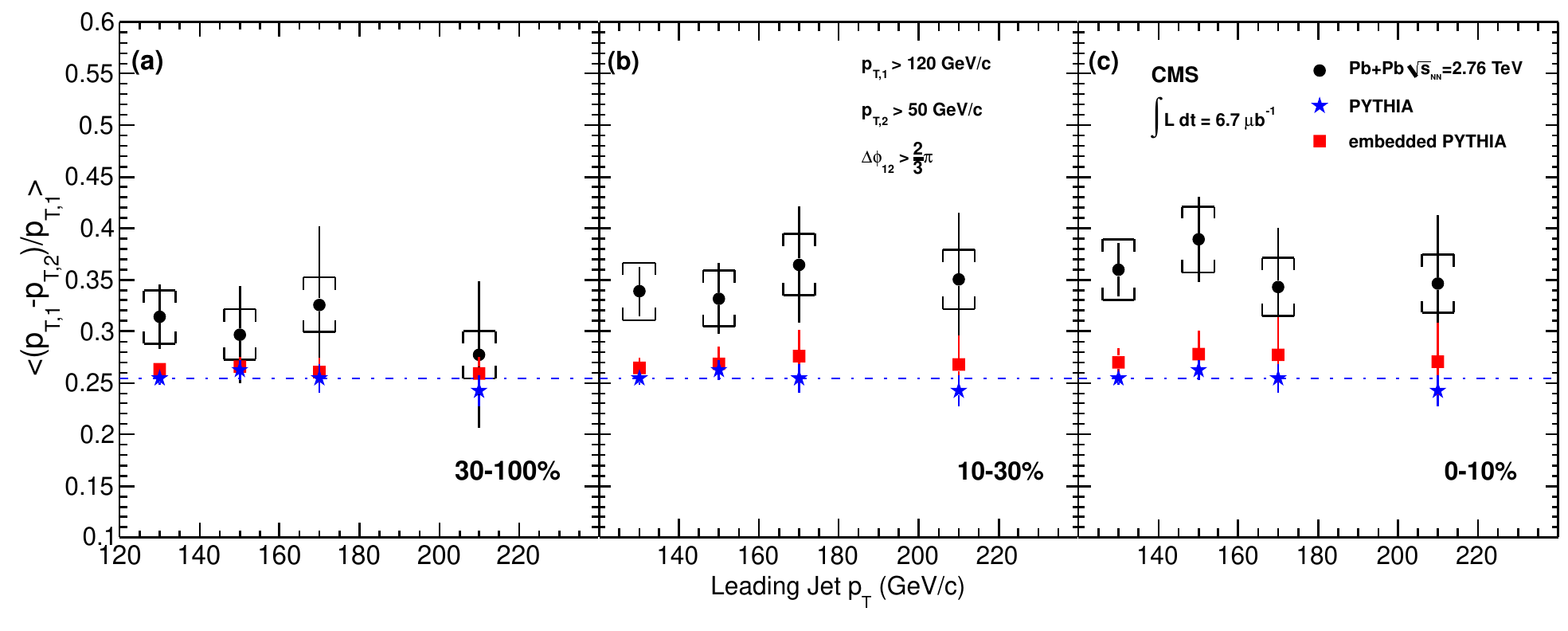}
\end{center}\vspace*{-2pc}
\caption[]{CMS measurement~\cite{CMSPRC} of $\mean{1-\hat{p}_{t,2}/\hat{p}_{t,1}}$, the fractional jet imbalance, for 3 centralities in Pb+Pb collisions at $\sqrt{s_{NN}}=2.76$ TeV, compared to PYTHIA simulations for p-p collisions.}
\label{fig:CMSAJ2011}
\end{figure}

\begin{figure}[!h]
\begin{center}
\includegraphics[width=0.84\textwidth]{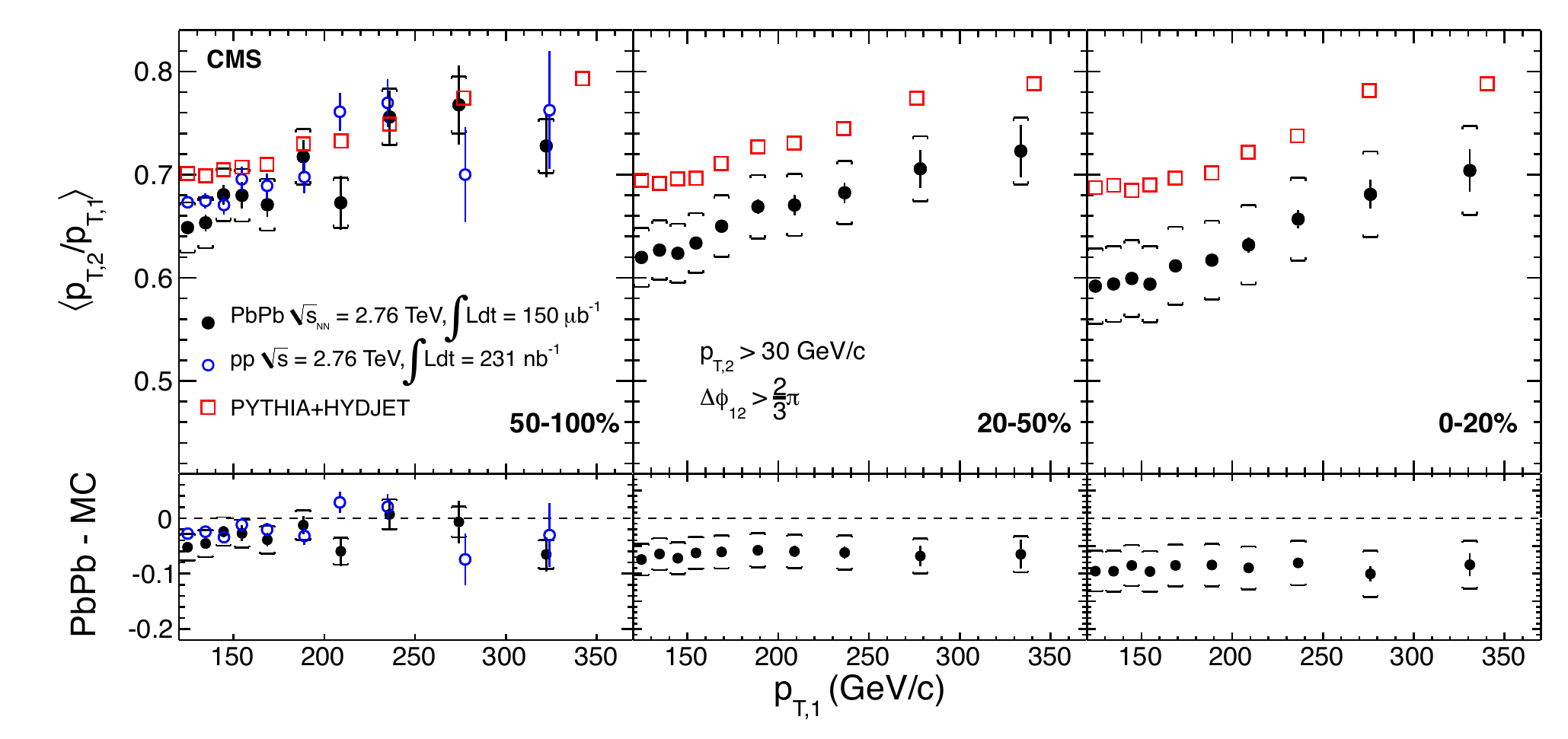}
\end{center}\vspace*{-2pc}
\caption[]{CMS~\cite{CMS-dijet-PLB712} measurements of average di-jet transverse momentum ratio,  $\hat{x}_h=\hat{p}_{T,2}/\hat{p}_{T,1}$, as a function of leading jet $\hat{p}_{T,1}$ at $\sqrt{s_{NN}}=2.76$ TeV in p-p collisions and for 3 centralities in Pb+Pb collisions, as well as simulated p-p di-jets embedded in heavy ion events.}   
\label{fig:CMSdijet2012}
\end{figure}

The fractional jet imbalance was also measured directly with reconstructed di-jets by the CMS collaboration at the LHC in Pb+Pb central collisions at $\sqrt{s_{\rm NN}}=2.76$ TeV (Fig.~\ref{fig:CMSAJ2011})~\cite{CMSPRC}; and with the large effect in p-p collisions corrected in the same way~\cite{MJT-Utrecht}, the results compared to PHENIX are shown in Fig.~\ref{fig:AuAupp79}c. New results this year by CMS (Fig.~\ref{fig:CMSdijet2012})~\cite{CMS-dijet-PLB712} confirm the  correction~\cite{MJT-Utrecht} and significantly extend and improve their previous measurement. 

The large difference in fractional jet imbalance between RHIC and LHC c.m. energies (Fig.~\ref{fig:AuAupp79}c) could be due to the difference in jet $\hat{p}_{T_t}$ between RHIC ($\sim 20$ GeV/c) and LHC ($\sim 200$ GeV/c), the difference in $n$ for the different $\sqrt{s_{NN}}$, or to a difference in the properties of the medium. In any case the strong $\hat{p}_T$ dependence of the fractional jet imbalance (apparent energy loss of a parton) also seems to disfavor purely radiative energy-loss in the \QGP~\cite{BSZ} and indicates that the details of energy loss in a \QGP\ remain to be understood.  Future measurements will need to sort out these issues by extending both the RHIC and LHC measurements to overlapping regions of $\hat{p}_T$.

\end{document}